\documentclass{article}

\usepackage[preprint]{neurips_2024}

\usepackage[utf8]{inputenc} 
\usepackage[T1]{fontenc}    
\usepackage{hyperref}       
\usepackage{url}            
\usepackage{booktabs}       
\usepackage{amsfonts}       
\usepackage{nicefrac}       
\usepackage{microtype}      
\usepackage{xcolor}         
\usepackage{graphicx}
\usepackage{amsmath}
\usepackage{subcaption}
\usepackage{multirow}

\title{A Knowledge-Inspired Hierarchical Physics-Informed Neural Network for Pipeline Hydraulic Transient Simulation}

\author{Jian Du$^1$\thanks{jiandu1997@163.com}, Haochong Li$^1$, Qi Liao$^1$\thanks{qliao@cup.edu.cn}, Jun Shen$^2$, Jianqin Zheng$^3$, Yongtu Liang$^1$\\
  $^1$China University of Petroleum-Beijing\\
  $^1$Fuxue Road No. 18, Changing District, Beijing 102249, PR China \\
  $^2$University of Wollongong \\
  $^2$Wollongong, NSW, Australia \\
  $^3$China Petroleum Planning and Engineering Institute \\
  $^3$Xinxi Road No 3, Haidian District, Beijing 100083, PR China \\
}

\begin{document}

\maketitle

\begin{abstract}
The high-pressure transportation process of pipeline necessitates an accurate hydraulic transient simulation tool to prevent slack line flow and over-pressure, which can endanger pipeline operations. However, current numerical solution methods often face difficulties in balancing computational efficiency and accuracy. Additionally, few studies attempt to reform physics-informed learning architecture for pipeline transient simulation with magnitude different in outputs and imbalanced gradient in loss function. To address these challenges, a Knowledge-Inspired Hierarchical Physics-Informed Neural Network is proposed for hydraulic transient simulation of multi-product pipelines. The proposed model integrates governing equations, boundary conditions, and initial conditions into the training process to ensure consistency with physical laws. Furthermore, magnitude conversion of outputs and equivalent conversion of governing equations are implemented to enhance the training performance of the neural network. To further address the imbalanced gradient of multiple loss terms with fixed weights, a hierarchical training strategy is designed. Numerical simulations demonstrate that the proposed model outperforms state-of-the-art models and can still produce accurate simulation results under complex hydraulic transient conditions, with mean absolute percentage errors reduced by 87.8\% and 92.7 \% in pressure prediction. Thus, the proposed model can conduct accurate and effective hydraulic transient analysis, ensuring the safe operation of pipelines.
\end{abstract}

\section{Introduction}

\subsection{Background}

Refined products are indispensable and strategic energy resources that significantly impact the stable development of the global economy and society (\cite{DU2023127452}). In 2023, refined product consumption reached 403 million tons, according to data released by the National Bureau of Statistics of China. Pipeline transportation, known for its cost-effectiveness and reliability, is the predominant method for delivering refined oil from refineries to market depots (\cite{LU2023162386}). The high-pressure transportation process and the inevitable aging of pipelines impose stringent requirements on safe operation (\cite{ZHENG2022125025}). Specifically, product vaporization and slack line flow at high elevation regions increase water hammer pressure (\cite{10.1115/IPC2004-0597}), posing significant risks to pipeline safety (\cite{PAN2024112833}). Therefore, developing accurate flow simulation techniques to estimate pressure and flowrate along the entire pipeline is essential (\cite{YIN2023120615}).

\begin{figure}
    \centering
    \includegraphics[width=0.75\linewidth]{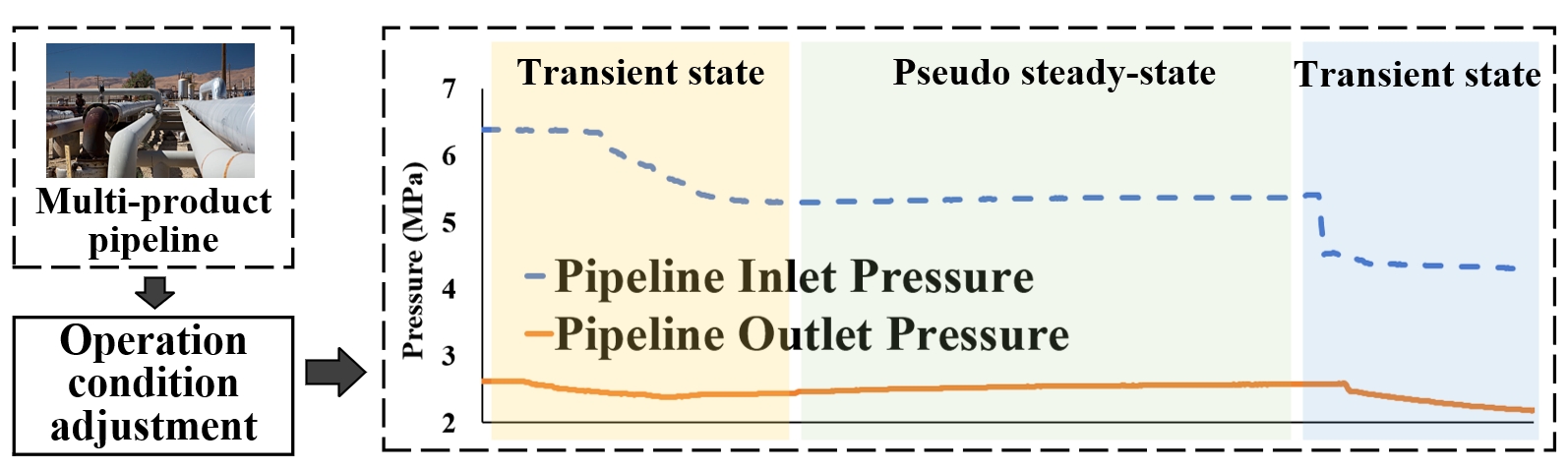}
    \caption{Schematic diagram of variations in pipeline pressure and flowrate}
    \label{fig:1}
\end{figure}

Flow simulation of multi-product pipelines involves pseudo unsteady-state simulation and transient simulation, as depicted in Figure \ref{fig:1}. Due to variable downstream market demand, diversification of oil types, and the need for energy conservation and consumption reduction, pipeline operating conditions frequently change (\cite{ZHENG2021510}). Consequently, during the transient process of tight line transportation, pressure and flowrate may fluctuate violently, leading to slack line flow at high elevations and abrupt pressure increases at low elevations. As pipelines age and corrode, exceeding the maximum allowable operating pressure (MAOP) can result in major safety incidents, such as bursts and explosions. Thus, with the increasing demand for pipeline operational safety, rapid and accurate flow simulation of transient processes is urgently needed (\cite{ZHENG2022125025}).

\subsection{Literature review}

Recently, some scholars conducted research to estimate the pipeline transient flow parameters. The methods employed for transient flow simulation can be categorized into two main types: model-driven methods and data-driven methods. Model-driven methods describe the hydraulic transient process of pipelines using various nonlinear partial differential equations (PDEs), continuity equations, and momentum equations, which encapsulate the implicit physical background of the process. Numerical discrete methods are then utilized to estimate pressure and flowrate based on operational data from the pipeline inlet and outlet.

To mitigate the interference of measurement noise in operational data, several studies have employed filter-based transient simulation methods (\cite{DELGADOAGUINAGA2021104888,https://doi.org/10.1002/asjc.1362,HE202019}). These methods estimate physical quantities based on model parameters. However, transient flow parameters are uncertain due to variations in wall roughness, product properties, and other factors. Acquiring reliable model parameters through extensive expert reasoning and analysis of large historical datasets is both time-consuming and labor-intensive, reducing computational flexibility. Consequently, some studies have incorporated optimization methods into pipeline transient simulation models to minimize deviations (\cite{HE202019,MA2022595}). The deviation calibration process, essentially an inverse problem, requires numerous evaluations and solutions of PDEs, posing a significant challenge for efficient transient analysis due to its computational expense (\cite{ZHANG2023106073}).

With the rapid advancements in data storage and computing capacity (\cite{DU2023128810}), data-driven models are designed to achieve better accuracy transient simulation in pipelines (\cite{YONG2020102024,HE202019,RIDOLFI2014648}). However, current data-driven methods fail to monitor flow parameters along the entire pipeline. More importantly, since the training of data-driven models relies solely on available data, the results may lack explainability from a physics perspective due to the omission of scientific theory of pipeline transient flow.

In 2019, a new paradigm emerged for addressing scientific problems by incorporating scientific theories into deep neural networks (DNN), known as the physics-informed neural network (PINN) is proposed by \cite{RAISSI2019686}. PINNs can simultaneously solve inverse and forward problems without requiring a fully known governing equation, eliminating the need for PDE evaluation and solution (\cite{chen2021physics}). \cite{YE2022118828} designed PINN-based architecture for hydraulic transient analysis in water pipeline. However, the imbalanced gradient descent that is caused by significant magnitude difference in outputs and fixed-weight multiple loss terms haven't been resolved for better accurate state estimation of liquid pipelines.

To fill the gap in the current literature, an innovative physics-informed deep learning method is proposed in this paper. Initially, magnitude conversion on the outputs is performed, and the PDEs are equivalently transformed to obtain output derivatives through automated differentiation (AD). Subsequently, the physics laws of the transient flow process are utilized to derive the coupling loss function. To tackle the performance deficiency with fixed weight loss function, a hierarchical training strategy is designed to decompose the training of PINN into  multiple sub-processes. To start with, the proposed model, named Knowledge-Inspired Hierarchical Physics-Informed Neural Network (KIH-PINN), offers a promising and more universal physics-informed learning framework for flow simulation in multi-product pipeline and other liquid pipelines, with notable transferability and high computational flexibility.

The contributions of this work can be concluded as follows:
\begin{enumerate}
    \item Governing equations, boundary condition, and initial condition are mathematically encoded to penalize the violation of physics principles during the training of neural network, intending to enforce physically explainable flow simulation with highly computational flexibility and efficiency.
    \item  The training process of PINNs are decomposed into multiple sub-processes concerning the penalty terms and governing equations are converted equivalently to prevent the deficient gradient descent performance with fixed weight loss function and improve the convergence effect.
    \item Results from simulated pipeline cases indicates that the proposed model captures more accurate dynamic hydraulic characteristics than conventional PINN and DNN, with RMSE reduced by 42 \% and 91 \%.
\end{enumerate}

The remainder of this paper is organized as follows. Section \ref{section2} delves into the mathematical basis of the transient flow process in multi-product pipelines and introduces the innovative physics-informed modeling framework. Section \ref{headings} presents a comprehensive case study and discussion on a simulated pipeline. Finally, Section \ref{conclusion} summarizes the conclusions of this study.

\section{Methodology}
\label{section2}
\subsection{Problem description}

In a one-dimensional pipe with sightly compressible fluid inside, the continuity and the momentum equations can be expressed as follows:
\begin{equation}\label{eqn-1}
    \frac{\partial v}{\partial t}+v\frac{\partial v}{\partial x}+g\frac{\partial h}{\partial x}+f\frac{v\left|v\right|}{2D}=0\
\end{equation}
\begin{equation}\label{eqn-2}
    \frac{\partial h}{\partial t}+\nu\frac{\partial h}{\partial x}+\frac{a^2}{g}\frac{\partial\nu}{\partial x}=0\
\end{equation}
\begin{equation}\label{eqn-4}
a=\sqrt{\frac{K/\rho}{1+\frac{K}{E}\frac{D}{\delta}C_1}}
\end{equation}

where \textit{h} is the head, m. \textit{v} is the velocity, m/s. $\rho$ is the fluid density, kg/m\textsuperscript{3}. \textit{x} is the distance along the pipeline, m. \textit{t} is the time, s. \textit{a} is the wave speed, m/s. \textit{g} is the gravitational acceleration, m2/s. \textit{A} is the cross-sectional area of the pipeline, m2. \textit{D} is the pipeline diameter, m. \textit{f} is the Darcy–Weisbach friction factor. \textit{K} is the volume elasticity modulus of fluid, Pa. \textit{E} is the elasticity modulus of pipeline, Pa.  $\delta$  is the pipe wall thickness, m.  $C_{1}$  is the pipeline constraint coefficient.

Given the pipeline parameters \(\beta=\{a,D,L,f\}\) (L is the pipeline length), estimating the transient flow state involves determining the state vector by solving the governing equations (\ref{eqn-1} and \ref{eqn-2}) and the conversion formula (\( P=\frac{\rho gh}{10^6}\), \(q=A v\), where \textit{P} is the pressure, MPa. \textit{q} is the flowrate per unit volume cross-section per second, m\textsuperscript{3}/s).

\begin{figure}
    \centering
    \includegraphics[width=0.8\linewidth]{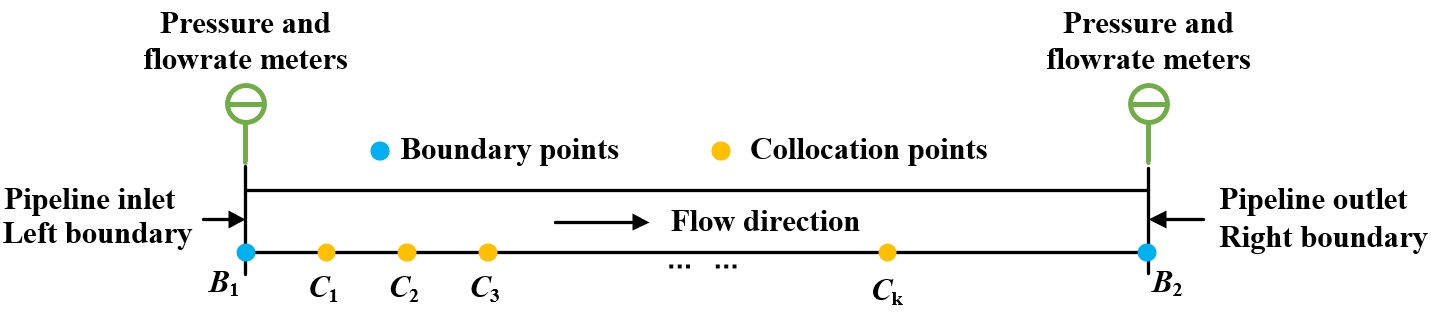}
    \caption{Pipeline schematic diagram in practical}
    \label{fig:2}
\end{figure}

Typically, boundary condition can be determined by the operation data from high-frequency sensors, as shown in Figure \ref{fig:2}{}. The transient flow process generally starts with a pseudo-steady state. Consequently, the head at the collocation points in the initial state of transient flow can be determined using Darcy’s friction law, while the velocity at these points remains nearly constant (\cite{chen2021physics}). Collocation points are predefined locations where hydraulic states adhere to the governing equations of transient pipeline flow.

\subsection{Knowledge-Inspired Hierarchical Physics-Informed Neural Network}
\subsubsection{A Knowledge-Inspired Physics-Informed Neural Network}

\begin{figure}[h]
	
	\begin{minipage}{0.32\linewidth}
		\vspace{3pt}
		\centerline{\includegraphics[width=0.8\textwidth]{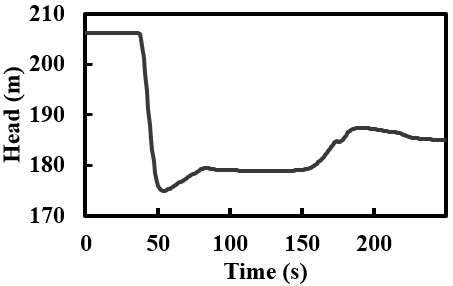}}
		\centerline{(a) Head}
	\end{minipage}
	\begin{minipage}{0.32\linewidth}
		\vspace{3pt}
		\centerline{\includegraphics[width=0.8\textwidth]{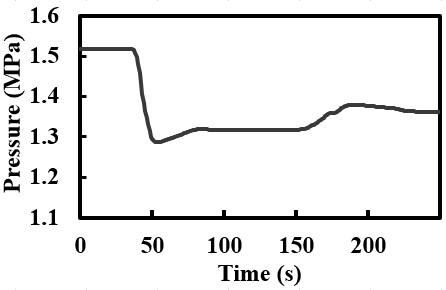}}
	 
		\centerline{(b) Pressure}
	\end{minipage}
	\begin{minipage}{0.32\linewidth}
		\vspace{3pt}
		\centerline{\includegraphics[width=0.8\textwidth]{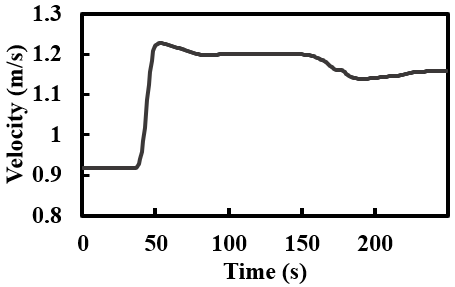}}
	 
		\centerline{(c) Velocity}
	\end{minipage}
 
	\caption{Comparison of magnitude between velocity and head}
	\label{fig:3}
\end{figure}

To approximate the nonlinear relationship between spatial-temporal coordinates and flow parameters (head and velocity), a DNN is constructed. The conventional neural network tunes the weights and biases using the mean squared errors (MSE) between predicted results and observed values (\cite{10622103}):

However, due to the significant magnitude difference, imbalanced gradients exist in the loss terms of the predicted head and velocity, as shown in Figure \ref{fig:3}. This can negatively affect the convergence of the predicted velocity, thereby impacting the accuracy of hydraulic state estimation.

\begin{figure}
    \centering
    \includegraphics[width=0.9\linewidth]{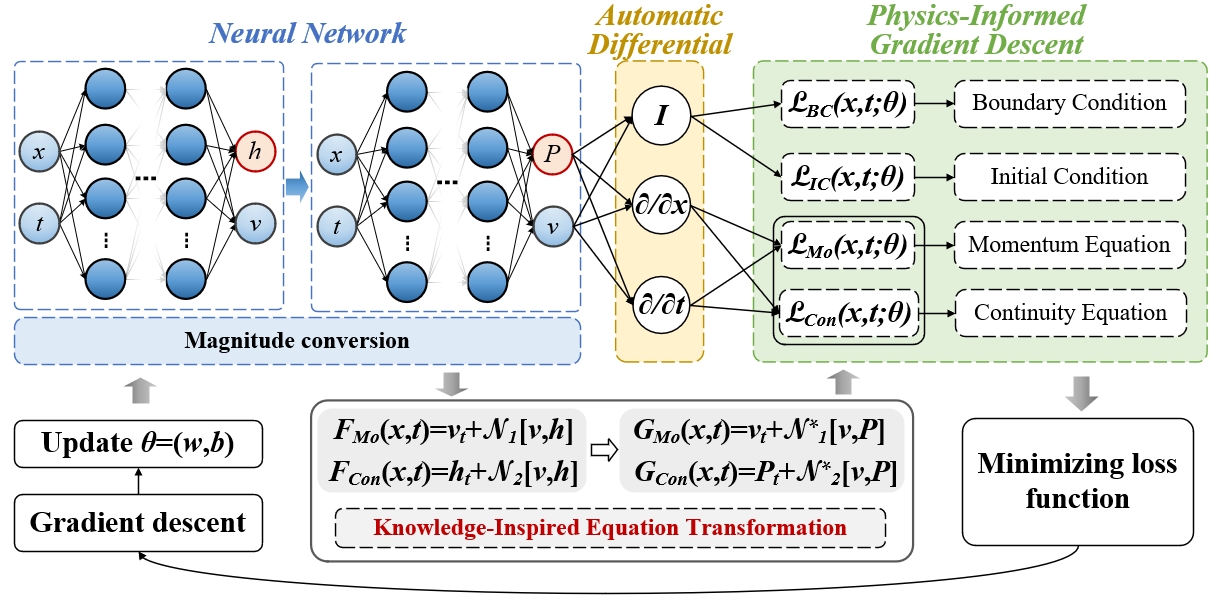}
    \caption{The framework of the proposed knowledge-inspired physics-informed neural network}
    \label{fig:4}
\end{figure}

Notably, the pressure has a similar magnitude to the velocity. Therefore, the neural network outputs are converted to pressure and velocity, as shown in Figure \ref{fig:4}. For desirable performance, training a DNN requires a substantial amount of observed data. However, in practice, only boundary data are available for pipeline transient simulation. This implies that the neural network must estimate complex, unknown hydraulic states at numerous collocation points, relying solely on boundary data. This challenging issue places stringent demands on the model extrapolation capability (\cite{PAN2024104658}). To ensure the outputs adhere to the physical laws of pipeline transient flow, the scientific theories should be integrated into the neural network. 

As depicted in Figure \ref{fig:4}, It is important to equivalently convert the governing equations to maintain the identity of the physical quantities. The continuity and momentum equations which are converted based on  Eq (\ref{eqn-1}) and (\ref{eqn-2}) are shown in Eq (\ref{eqn-7}) and (\ref{eqn-8}):

\begin{equation}\label{eqn-7}
    \frac{\rho g}{10^{6}}\frac{\partial v}{\partial t}+\frac{\rho vg}{10^{6}}\frac{\partial v}{\partial x}+g\frac{\partial P}{\partial x}+f \frac{\rho g}{10^{6}}\frac{v\left|v\right|}{2D}=0
\end{equation}
\begin{equation}\label{eqn-8}
    \frac{\partial P}{\partial t}+\nu \frac{\partial P}{\partial x}+\frac{\rho a^{2}}{10^{6}}\frac{\partial v}{\partial x}=0
\end{equation}

During the training of neural network, the predicted pressure and velocity can be approximated as $NN_{_P}$ and $NN_{_v}$. In this way, the residual of continuity and momentum equations can be expressed mathematically as follows:

\begin{equation}\label{eqn-9}
    G_{Mo}=\frac{\rho g}{10^6}\frac{\partial NN_\nu}{\partial t}+\frac{\rho gNN_\nu}{10^6}\frac{\partial NN_\nu}{\partial x}+g\frac{\partial NN_P}{\partial x}+f\frac{\rho g}{10^6}\frac{NN_\nu\left|NN_\nu\right|}{2D}
\end{equation}
\begin{equation}\label{eqn-10}
    G_{Con}=\frac{\partial NN_{P}}{\partial t}+\nu\frac{\partial NN_{P}}{\partial x}+\frac{\rho a^{2}}{10^{6}}\frac{\partial NN_{\nu}}{\partial x}
\end{equation}

where $G_{Mo}$ and $G_{Con}$ are the residual terms of continuity and momentum equations. Note that highly precise model parameters are not required to derive these residual terms. Smaller residual values indicate higher compliance. Consequently, two loss terms can be constructed by computing the partial derivatives using the AD algorithm to penalize violations of the continuity and momentum equations, as shown in Eq (\ref{eqn-11}) and (\ref{eqn-12}):

\begin{equation}\label{eqn-11}
    \mathcal{L}_{Mo}=\frac{1}{N_{f}}\sum_{i=1}^{N_{f}}\left|G_{Mo}\left(x_{f}^{i},t_{f}^{i}\right)\right|^{2}
\end{equation}
\begin{equation}\label{eqn-12}
 \mathcal{L}_{Con}=\frac1{N_f}\sum_{i=1}^{N_f}\left|G_{Con}\left(x_f^i,t_f^i\right)\right|^2
\end{equation}

where $\begin{Bmatrix}x_{f}^{i},t_{f}^{i}\end{Bmatrix}$ denotes the spatial-temporal coordinates at collocation points. $N_{_f}$ is the number of collocation points. Note that the estimated states from the neural network at collocation points must adhere to the continuity and momentum equations, and evaluating their penalty terms does not require the true target values of pressure and velocity. To ensure the results acquired by the neural network adhere to boundary and initial conditions, loss terms ought to be reflected in the training process to penalize the violation, as depicted in Eq (\ref{eqn-15}) and (\ref{eqn-16}):

\begin{equation}\label{eqn-15}
\mathcal{L}_{BC}=\frac{1}{N_{BC}}\sum_{i=1}^{N_{BC}}\left|\frac{\left[NN_{P}\left(x_{BC}^{i},t^{i};\theta\right)-P_{BC}^{i}\right]+\left[NN_{\nu}\left(x_{BC}^{i},t^{i};\theta\right)-\nu_{BC}^{i}\right]}{2}\right|^{2}
\end{equation}

\begin{equation}\label{eqn-16}
\mathcal{L}_{IC}=\frac1{N_{IC}}\sum_{i=1}^{N_{IC}}\left|\frac{\left[NN_P\left(x^i,t_{IC}^i;\theta\right)-P_{IC}^i\right]+\left[NN_\nu\left(x^i,t_{IC}^i;\theta\right)-\nu_{IC}^i\right]}2\right|^2
\end{equation}

where $x_{BC}$ is the spatial coordinates at the boundary, and $t_{IC}$ is the temporal coordinate at the initial state. $P_{BC}$ and $v_{BC}$ represent the observed pressure and velocity at the boundary. $P_{IC}$ and $v_{IC}$ denote the observed pressure and velocity at the initial state.  $N_{BC}$ and $N_{IC}$ are the number of samples at boundary and initial conditions. In this way, a coupling loss function can be acquired by summing up all the loss terms as follows:

\begin{equation}\label{eqn-17}
\mathcal{L}\left(\theta\right)=\lambda_{BC}\mathcal{L}_{BC}+\lambda_{IC}\mathcal{L}_{IC}+\lambda_{Con}\mathcal{L}_{Con}+\lambda_{Mo}\mathcal{L}_{Mo}
\end{equation}

where $\{\lambda_{_{BC}},\lambda_{_{IC}},\lambda_{_{Con}},\lambda_{_{Mo}}\}$ is the hyper-parameters, which can be adjusted during the training process to balance the weights of different loss terms. By minimizing the coupling loss function through the gradient descent process, the hydraulic states estimated by neural network are expected to align with the physical laws governing pipeline transient flow.

\subsubsection{Hierarchical Training Strategy}

During neural network training, the gradient descent of different loss terms is influenced by the order of magnitude differences in residual values, which vary dynamically. Numerical experiments indicate that using a fixed-weight coupling loss function may lead to a local optimal solution \cite{XIANG202211}. To tackle this challenge, the training of multiple loss terms is decomposed into several stages to achieve hierarchical convergence for each term. The proposed training strategy aims to develop a more reproducible gradient-balance method for training PINNs in engineering practice, while simplifying the mathematics. The execution process of this training strategy is as follows:

\paragraph{Stage One}

Given that the two-dimensional variables of the neural network must adhere to boundary conditions, the first stage involves reorganizing the residual terms of the boundary conditions for model training. As illustrated in Eq (\ref{eqn-15}) and (\ref{eqn-16}),  loss terms that are applied to measure the compliance of the predicted flow parameters with boundary conditions in this stage can be expressed as:

\begin{equation}\label{eqn-20}
\mathcal{L}_{\mathrm{Stage~one}}(\theta)=\frac{1}{N_{BC}}\sum_{i=1}^{N_{BC}}\left|F_{BC}^{\nu}(x_{BC}^{i},t^{i})\right|^{2}+\frac{1}{N_{BC}}\sum_{i=1}^{N_{BC}}\left|F_{BC}^{P}(x_{BC}^{i},t^{i})\right|^{2}
\end{equation}

With the guidance of the loss function in Eq (\ref{eqn-20}), the gradient descent process implemented to tune the parameters, as shown in Eq (\ref{eqn-21}):
\begin{equation}\label{eqn-21}
\theta^{(1)}=\arg\min\left(\mathcal{L}_{\text{Stage one}}(\theta)\right)
\end{equation}

where $\theta^{(1)}$ is the acquired network parameters after the training is finished. In the first stage, the network parameters are initialized randomly to find the optimal weights.

\paragraph{Stage Two} 
Before the training in stage two, the network parameters in stage one are applied to initialize the parameters in neural network in stage two. The residual errors used to quantify the compliance of predicted flow parameters with initial conditions in stage two can be represented as:

\begin{equation}\label{eqn-24}
L_{Stagetwo}(\theta)=\frac{1}{N_{IC}}\sum_{i=1}^{N_{IC}}\left|F_{IC}^P(x^i,t_{IC}^i)\right|^2+\frac{1}{N_{IC}}\sum_{i=1}^{N_{IC}}\left|F_{IC}^{\nu}(x^{i},t_{IC}^{i})\right|^{2}
\end{equation}
By minimizing the loss function in Eq (\ref{eqn-24}), the gradient descent process is carried out to train the network parameters, as depicted in Eq (\ref{eqn-25}):

\begin{equation}\label{eqn-25}
\theta^{(2)}=\arg\min_{\theta}\left(\mathcal{L}_{Stage two}\left(\theta\right)\right)\quad\mathrm{with} \theta=\theta^{(1)}
\end{equation}
where $\theta^{(2)}$ is the acquired network parameters after the training process in stage two is finished. In that way, the admissible solution space of neural network has been narrowed to honor the boundary and initial conditions.

\paragraph{Stage Three} 

In stage three, the coupling loss function, which consists of multiple residual terms related to governing equations, boundary conditions, and initial conditions, is designed to train the network parameters. Based on the definitions from Eq (\ref{eqn-11}) to (\ref{eqn-17}), the loss function in this stage can be expressed as follows:

\begin{equation}\label{eqn-26}
L_{Stage tlree}\left(\theta\right)=\lambda_{BC}L_{BC}+\lambda_{IC}L_{IC}+\lambda_{Con}L_{Con}+\lambda_{Mo}L_{Mo}
\end{equation}

To start with, the gradient descent process in stage three is implemented to acquire the weights and biases, as depicted in Eq (31):

\begin{equation}\label{eqn-27}
\theta^{(3)}=\arg\min_{\theta}\left(\mathcal{L}_{Stage three}\left(\theta\right)\right)\quad\mathrm{with} \theta=\theta^{(2)}
\end{equation}

where $\theta^{(3)}$ denotes the trained network parameters in stage three. Upon completing the third stage of training, the predicted results of neural network fall within the solution space that align closely with the scientific principles of transient flow. The Adam optimizer is employed in all three stages to optimize the network parameters (\cite{DU2022124689}).

\section{Case studies with experiments}
\label{headings}

\subsection{Experiment setting}

\begin{figure}
    \centering
    \includegraphics[width=0.8\linewidth]{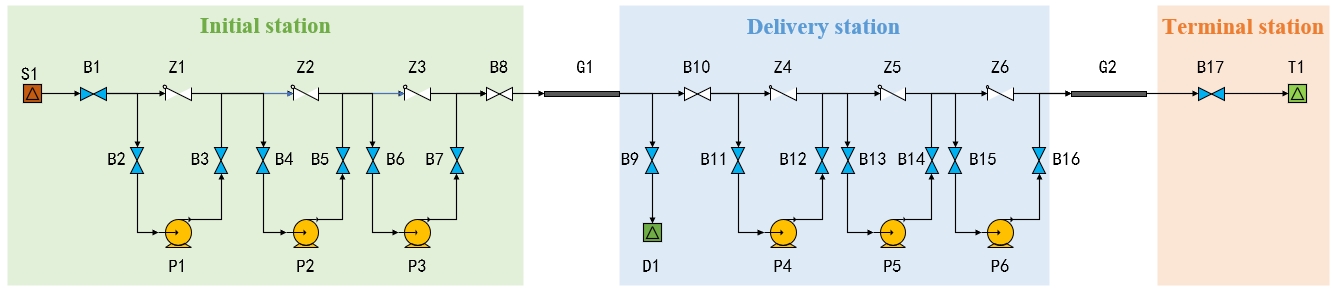}
    \caption{The diagrammatic sketch of the SPS simulation model of a pipeline system }
    \label{fig:5}
\end{figure}

As shown in Figure \ref{fig:5}, a simulated multi-product pipeline system, which consists of an initial station, a delivery station, and a terminal station, is constructed using SPS software (\cite{WANG2022107733}) to obtain pressure and flowrate along the pipeline. The simulation model of the multi-product pipeline system includes a supply point (S1) at the initial station, six oil pumps (P1~P6), two long-distance pipelines (G1, G2), seventeen gate valves (B1$ \sim $B17), six check valves (Z1$ \sim $Z3), a delivery point (D1), and an oil depot (T1) at the terminal station. The same parameter setting as in \cite{WANG2022107733} is used, with the pipeline lengths of S1-D1 and D1-T1 being 25 km each. The time step of the sampling data from the simulation model is determined as  =0.5 s. Points are evenly distributed at 1 km intervals, resulting in 50 points. In this paper, all these 48 collocation points are used to impose governing equations and initial condition to the neural network. Two boundary points at pipeline inlet and outlet are applied to impose boundary condition.

To validate the generality of KIH-PINN, a typical operating condition of a multi-product pipeline was simulated to generate experimental data. Table \ref{tab:1} presents the properties of the oil products and the boundary conditions for the experimental operation.

\begin{table}[]
\centering
\caption{The properties of oil products and boundary condition of different operation conditions}
\label{tab:1}
\begin{tabular}{ccccc}
\hline
\multicolumn{3}{c}{\begin{tabular}[c]{@{}c@{}}Product \\ propertys\end{tabular}}                                                                                                                          & \multicolumn{2}{c}{Boundary condition}                                                                                                        \\ \hline
\begin{tabular}[c]{@{}c@{}}Density/\\ (kg/m\textsuperscript{3})\end{tabular} & \begin{tabular}[c]{@{}c@{}}Viscosity/\\ (m\textsuperscript{2} /s)\end{tabular} & \begin{tabular}[c]{@{}c@{}}Volume \\ Elasticity \\ Coefficient/Pa\end{tabular} & \begin{tabular}[c]{@{}c@{}}Starting-point \\ pressure/MPa\end{tabular} & \begin{tabular}[c]{@{}c@{}}End-point \\ flowrate/(m\textsuperscript{3}/h)\end{tabular} \\
850                                                        & 5.2×10-6                                                    & 1.5×109                                                                        & 1.48                                                                   & 154                                                                  \\ \hline
\end{tabular}
\end{table}

In this study, the DNN model and the PINN model used for hydraulic transient analysis in water pipeline from \cite{YE2022118828} are applied as comparative baseline models. Several metrics in \cite{FU2024106098} are used to evaluate the model performance, including Root Mean Square Error (RMSE), Mean absolute percentage error (MAPE), and R-Square (R\textsuperscript{2} ). Through trial and error, a neural network with ten hidden layers is employed, each containing 50 units. The batch size is set to 128, and Softplus is used as the activation function. The total number of iterations is 20,000, with an initial learning rate of 0.0001.

\subsection{Transient hydraulic analysis}

\begin{figure}[ht]
    \centering
    \begin{subfigure}[b]{0.21\textwidth}
        \includegraphics[width=\textwidth]{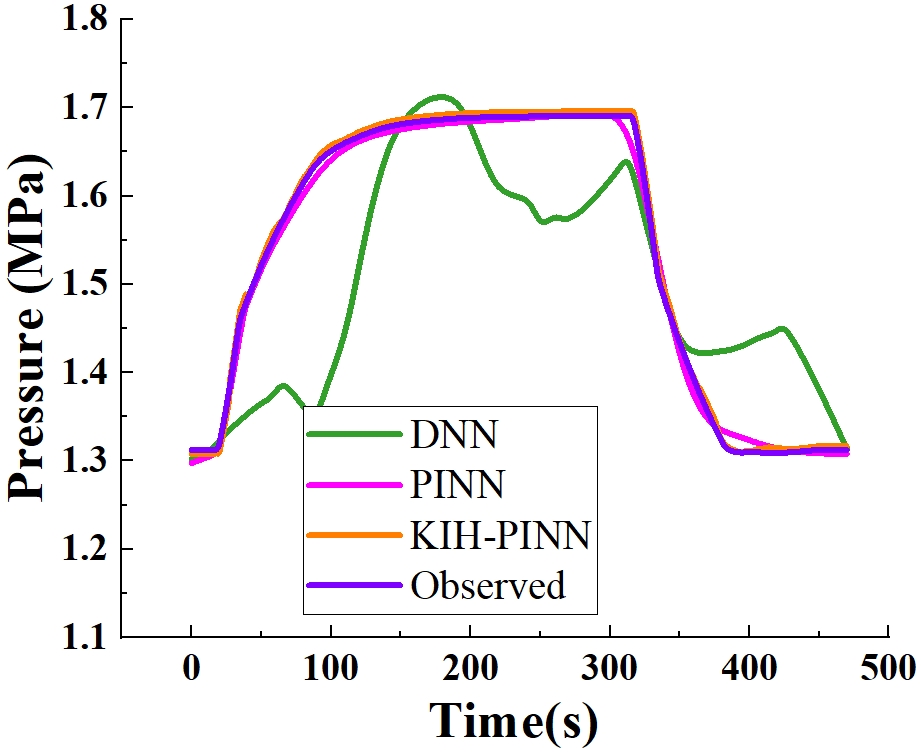}
        \caption{8 km}
    \end{subfigure}
    \hfill
    \begin{subfigure}[b]{0.21\textwidth}
        \includegraphics[width=\textwidth]{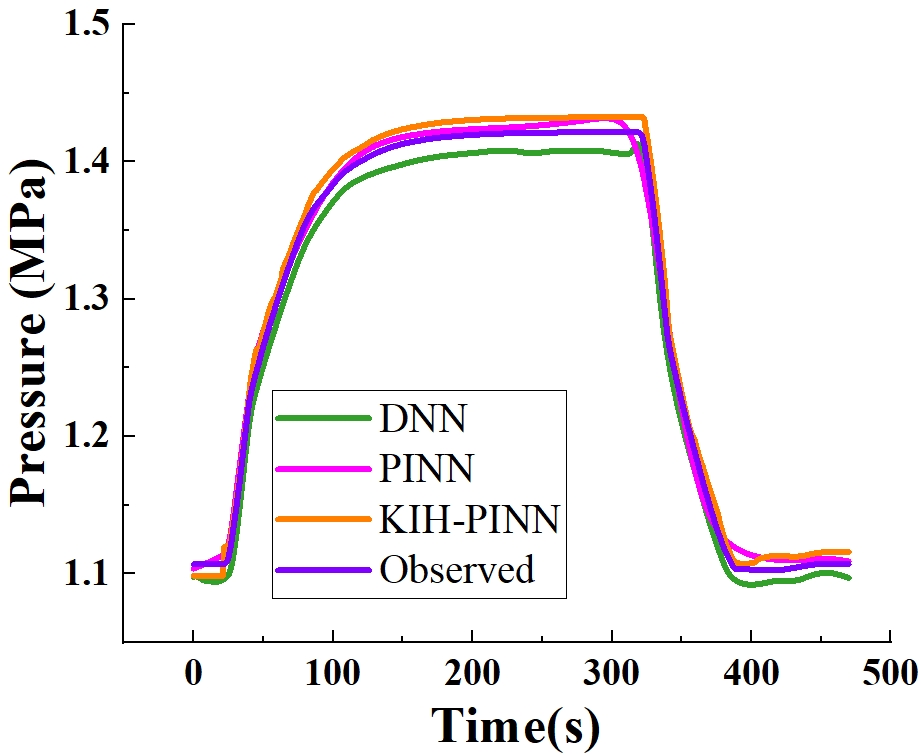}
        \caption{15 km}
    \end{subfigure}
    \hfill
    \begin{subfigure}[b]{0.21\textwidth}
        \includegraphics[width=\textwidth]{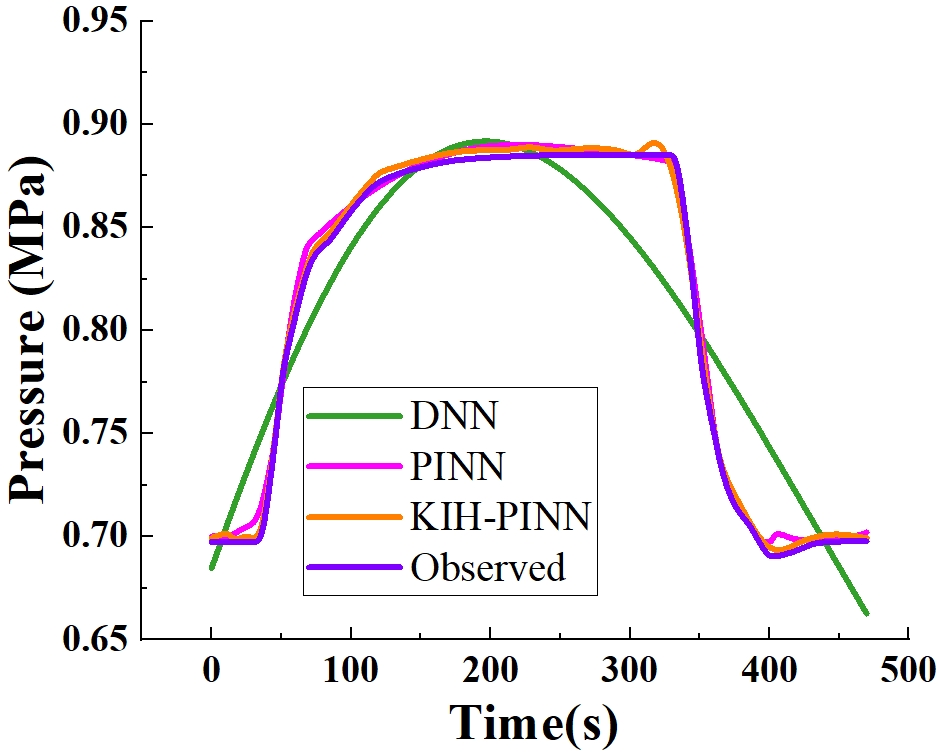}
        \caption{30 km}
    \end{subfigure}
    \hfill
    \begin{subfigure}[b]{0.21\textwidth}
        \includegraphics[width=\textwidth]{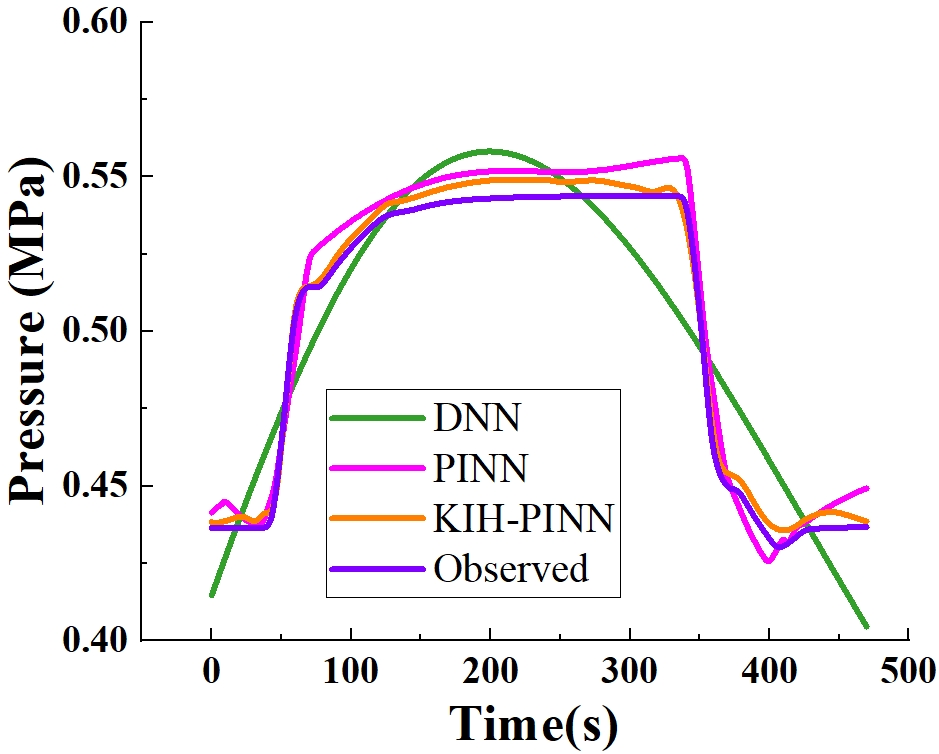}
        \caption{42 km}
    \end{subfigure}
\caption{The predicted pressure at different locations of various comparative models}
\label{fig:6}
\end{figure}

\begin{figure}[ht]
    \centering
    \begin{subfigure}[b]{0.21\textwidth}
        \includegraphics[width=\textwidth]{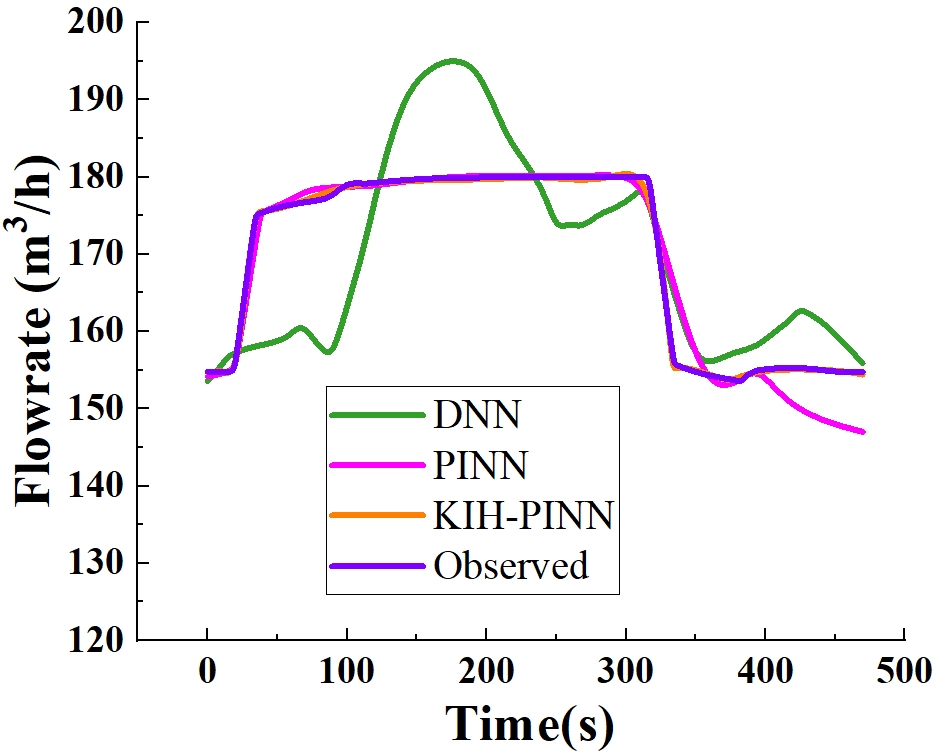}
        \caption{8 km}
    \end{subfigure}
    \hfill
    \begin{subfigure}[b]{0.21\textwidth}
        \includegraphics[width=\textwidth]{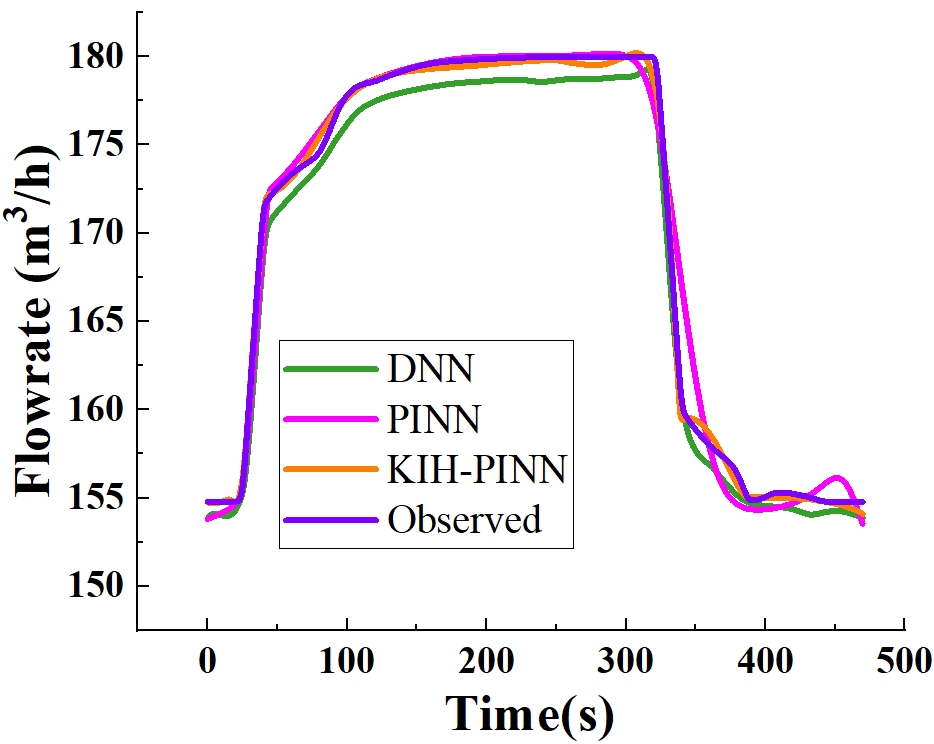}
        \caption{8 km}
    \end{subfigure}
    \hfill
    \begin{subfigure}[b]{0.21\textwidth}
        \includegraphics[width=\textwidth]{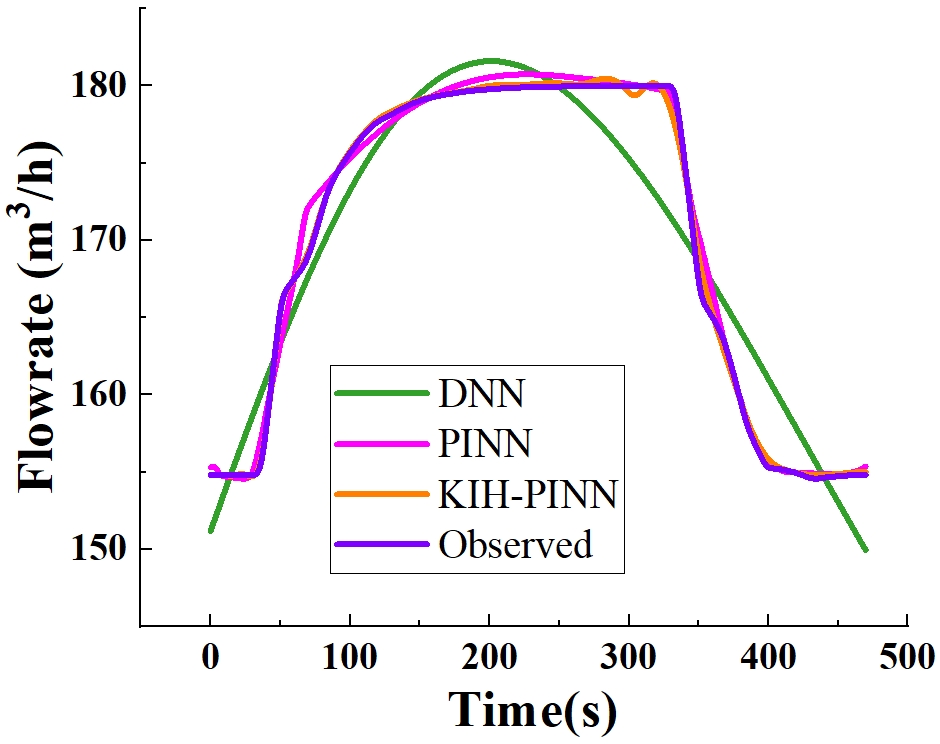}
        \caption{30 km}
    \end{subfigure}
    \hfill
    \begin{subfigure}[b]{0.21\textwidth}
        \includegraphics[width=\textwidth]{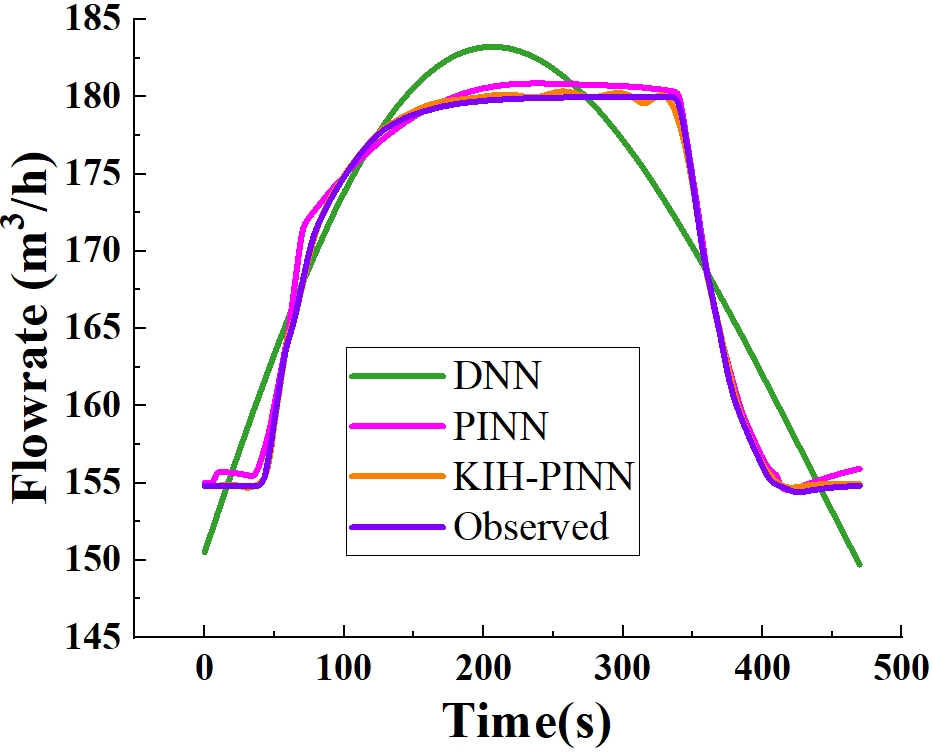}
        \caption{42 km}
    \end{subfigure}
\caption{The predicted flowrate at different locations of various comparative models}
\label{fig:7}
\end{figure}

The flow parameters at four random points predicted by different models under throughput increment and degradation condition are compared, as shown in Figure \ref{fig:6} and \ref{fig:7}. Overall, DNN performs the worst in transient simulation among comparative models. For example, the pressure and flowrate curves from DNN significantly deviate from the observed curves at 8 km. At 15 km, 30 km, and 42 km in pipeline G2, DNN only captures general trends, indicating further deterioration in prediction capacity. Conversely, PINN provides closer prediction curves than DNN by incorporating governing equations, boundary conditions, and initial conditions into the training process. Comparison of PINN to the proposed model shows that pressure and flowrate curves produced by the proposed model are the closest to the observed curves. This suggests that achieving a gradient-balanced training process through knowledge-inspired equation conversion and a hierarchical training strategy is crucial for developing a more accurate transient simulation model. Additionally, the proposed KIH-PINN achieves accuracy comparable to mainstream commercial simulation software, effectively enhancing computational flexibility and efficiency while extracting dynamic transient hydraulic behavior only relying on the monitoring data in boundaries.

\begin{figure}[ht]
    \centering
    \begin{subfigure}[b]{0.3\textwidth}
        \includegraphics[width=\textwidth]{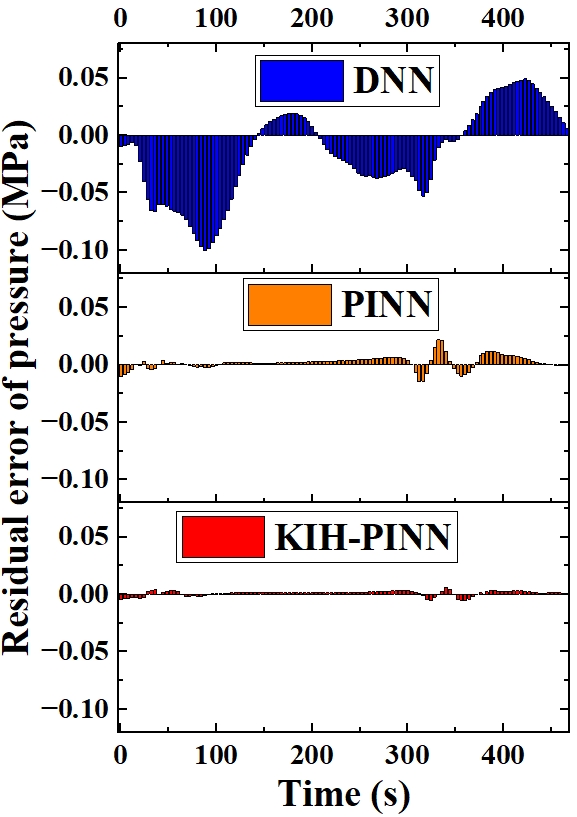}
        \caption{Pipeline G1}
    \end{subfigure}
    \hfill
    \begin{subfigure}[b]{0.3\textwidth}
        \includegraphics[width=\textwidth]{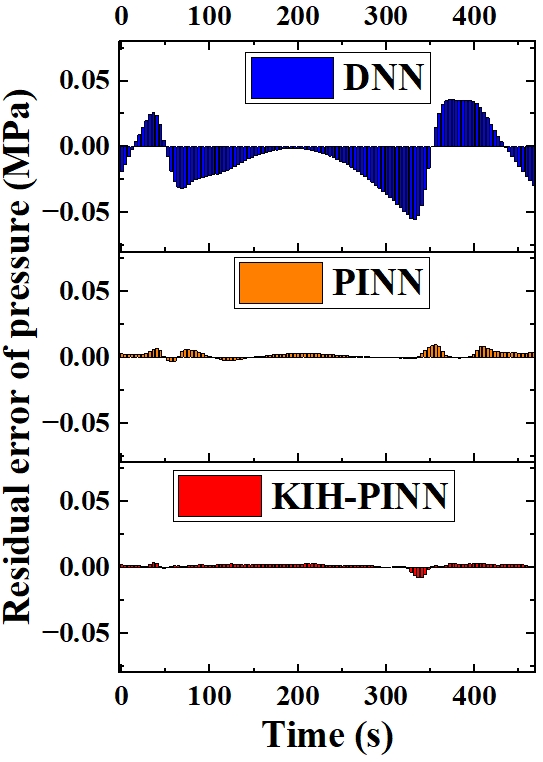}
        \caption{Pipeline G2}
    \end{subfigure}
\caption{The residual errors at each time step of different models}
\label{fig:8}
\end{figure}

To provide a more intuitive comparison and evaluation of prediction accuracy, the average residuals between the predicted pressure from various models and the actual values at all 48 non-detection points were calculated, as illustrated in Figure \ref{fig:8}. Overall, the residuals for the DNN model, represented by the blue bar, exhibit the largest area, indicating the worst predictive accuracy compared to other models. Additionally, these blue bars show significant peaks and troughs, highlighting the deficient ability of DNN model to capture hydraulic dynamic behavior. Conversely, the PINN model achieves a smaller bar area than the DNN by incorporating physical laws into forward and backward propagation. This demonstrates that integrating scientific theories and available data is essential for enhancing simulation performance. Notably, the red area representing the prediction deviations of KIH-PINN is significantly smaller than that of other models. Throughout the entire flow period, the maximum residual error of pressure predicted by the proposed model is approximately 0.01 MPa, whereas the maximum residual error predicted by PINN is nearly 0.05 MPa.

To quantitatively compare these three models, the evaluation metrics for predicted pressure and flowrate are presented in Table \ref{tab:2}. Overall, DNN exhibits the poorest approximation ability, with the highest prediction errors and the lowest fitting coefficient. For flowrate prediction in pipeline G2, the DNN model even produces a negative R² value. The PINN model demonstrates better prediction performance than DNN, with lower prediction errors. Finally, the proposed KIH-PINN model achieves the most accurate and effective predictions among all comparative models, with RMSE and MAPE values of 0.004 MPa and 0.617\% for pressure prediction in pipeline G2, respectively

\begin{table}[]
\centering
\caption{The evaluation metrics of predicted pressure and flowrate}
\label{tab:2}
\begin{tabular}{cccccccc}
\hline
\multirow{2}{*}{Pipeline} & \multirow{2}{*}{Model} & \multicolumn{3}{c}{\begin{tabular}[c]{@{}c@{}}Assessment of \\ predicted pressure\end{tabular}}                 & \multicolumn{3}{c}{\begin{tabular}[c]{@{}c@{}}Assessment of \\ predicted flowrate\end{tabular}}                   \\ \cline{3-8} 
                          &                        & \begin{tabular}[c]{@{}c@{}}RMSE\\ /MPa\end{tabular} & \begin{tabular}[c]{@{}c@{}}MAPE\\ /\%\end{tabular} & R²    & \begin{tabular}[c]{@{}c@{}}RMSE\\ /MPa\end{tabular} & \begin{tabular}[c]{@{}c@{}}MAPE\\ /\%\end{tabular} & R²      \\ \hline
\multirow{3}{*}{G1}       & DNN                    & 0.072                                               & 3.636                                             & 0.941 & 8.590                                               & 3.246                                             & 0.679   \\
                          & PINN                   & 0.009                                               & 1.865                                             & 0.999 & 1.843                                               & 0.585                                             & 0.975   \\
                          & KIH-PINN               & 0.008                                               & 0.445                                             & 0.999 & 0.360                                               & 0.140                                             & 0.999   \\
\multirow{3}{*}{G2}       & DNN                    & 0.061                                               & 8.474                                             & 0.915 & 10.527                                              & 5.791                                             & -37.409 \\
                          & PINN                   & 0.015                                               & 4.625                                             & 0.962 & 5.821                                               & 2.660                                             & 0.732   \\
                          & KIH-PINN               & 0.004                                               & 0.617                                             & 1.000 & 0.335                                               & 0.124                                             & 0.999   \\ \hline
\end{tabular}
\end{table}

\section{Conclusion}
\label{conclusion}

In this paper, a Knowledge-Inspired Hierarchical Physics-Informed Neural Network (KIH-PINN) is designed for hydraulic transient simulation in a multi-product pipeline. The primary advantages and innovations of the proposed model are summarized as follows:
\begin{enumerate}
    \item By integrating scientific principles absent in the available measurement data into the training process of neural network, the model is driven to generate physically plausible results. This physics-informed deep learning model can accurately estimate hydraulic states even amidst complex variations of transient flow and inaccurate system parameters, achieving a reduction of RMSE by 90\% and 71\% for pressure prediction in simulation pipelines compared to DNN.
    \item By converting the magnitudes of the outputs to achieve an equivalent transformation of the governing equations and employing hierarchical training strategy, the proposed model attains more precise hydraulic simulation results and overcomes the performance deficiencies. This approach is anticipated to develop a mathematically simplified and easily reproducible gradient-balanced method for training PINNs in engineering practice, resulting in a 76\% and 87\% reduction in MAPE for pressure simulation compared to the PINN model.

\end{enumerate}
The proposed model opens numerous opportunities for pipeline operation monitoring, intelligent scheduling, and security assurance with high accuracy and flexibility. While it can simulate accurate hydraulic states even with unknown pipeline parameters, it cannot determine specific parameters for further hydraulic analysis and security assessment. Future work will focus on optimal parameter identification through inverse problem solutions.

\section{Acknowledgement}

This work was partially supported by the National Natural Science Foundation of China (52202405), the ARC Linkage Project LP230100083, and the Science Foundation of China University of Petroleum, Beijing (2462023BJRC026). The authors are grateful to all study participants.

{
\small

\bibliographystyle{plainnat}
\bibliography{cites/citations-20240911T060802,cites/ScienceDirect_citations_1726024464904,cites/ScienceDirect_citations_1726033663541,cites/ScienceDirect_citations_1726034205261,cites/ScienceDirect_citations_1726037282565,cites/ScienceDirect_citations_1726037369692,cites/ScienceDirect_citations_1726038633196,cites/ScienceDirect_citations_1726038652064,cites/pericles_1934609319,cites/ScienceDirect_citations_1726039221600, cites/ScienceDirect_citations_1726039313871, cites/ScienceDirect_citations_1726039334390, cites/ScienceDirect_citations_1726039358579,cites/ScienceDirect_citations_1726040734691,cites/ScienceDirect_citations_1726040751899,cites/ScienceDirect_citations_1726040768716,cites/17ref,ScienceDirect_citations_1726041226846,cites/19ref,cites/21ref,cites/22ref,cites/23ref,cites/24,cites/25,cites/26,cites/27,cites/28,cites/29}
}


\end{document}